\documentstyle[preprint,aps]{revtex}

\begin{document}

\preprint{FIMAT-6/95}

\draft

\title{Electronic states in graded-gap junctions with band inversion}

\author{Francisco Dom\'{\i}nguez-Adame}

\address{Departamento de F\'{\i}sica de Materiales,
Facultad de F\'{\i}sicas, Universidad Complutense,
E-28040 Madrid, Spain}

\maketitle

\begin{abstract}

We theoretically study electronic states in graded-gap junctions of
IV-VI compounds with band inversion.  Using a two-band model within the
${\bf k}\cdot{\bf p}$ approximation and assuming that the gap and the
gap centre present linear profiles, we demonstrate the existence of a
set of localized states along the growth direction with a discrete
energy spectrum.  The envelope functions are found to be combination of
harmonic oscillator eigenfunctions, and the corresponding energy levels
are proportional to the square root of the quantum number.  The level
spacing can be directly controlled by varying the structure thickness.

\end{abstract}


\narrowtext

Narrow-gap IV-VI compounds like Pb$_{1-x}$Sn$_x$Te and
Pb$_{1-x}$Sn$_x$Se present band inversion under compositional variation.
In a band-inverted heterojunction the fundamental gap has opposite signs
on each side \cite{Seeger}.  For instance, Pb$_{1-x}$Sn$_x$Se undergoes
band inversion as the Sn molar fraction is increased: At $x=0.14$ the
gap vanishes whereas the negative one reaches the magnitude of the PbSe
gap at $x=0.28$.  Recently, such heterojunctions have received much
attention because subbands of electron-like and hole-like localized
interface states are formed with energy lying within the fundamental gap
\cite{Korenman,Agassi1,Agassi2,PSS}, a particular feature not met in
most common III-V heterojunctions. In those compounds the electronic
states near the gap are properly described by means of a two-band model
using the effective ${\bf k}\cdot {\bf p}$ approximation
\cite{Korenman}, where quite strong coupling of host bands in the
semiconductor takes place.  The equation governing conduction- and
valence-band envelope-functions in a simple two-band model, neglecting
far-band corrections, is a Dirac-like equation.  In view of the analogy
existing between the two-band model and the Dirac equation, the exact
solution can be found since one can use elaborated techniques like those
related to supersymmetric quantum mechanics \cite{Pan}.

In this paper we exploit further such a formal similarity to demonstrate
for the first time the existence of a new type of localized electronic
states whose envelope functions resemble those of the harmonic
oscillator but, unlike the case of the well-known parabolic quantum
wells, there are no classical turning points.  In the present structure
the confining potential leading to localized states enters in the
equation of motion as a position-dependent mass term, whereas in the
case of parabolic quantum wells the conduction-band modulation appears
as an electrostatic-like interaction within the framework of a one-band
model Hamiltonian.  In our approach, we require a graded-gap structure
with band inversion.  An appropriate graded doping may create a
modulation of both conduction- and valence-band with linear profiles,
which manifest themselves through the occurrence of a linear scalar
potential in the Dirac-like Hamiltonian.  For our present purposes, we
then take advantage that there exists a number of linear potentials for
which the Dirac equation reduces to Schr\"odinger-like harmonic
oscillator equations \cite{EL}, and then exact solutions can be found
in a closed form.

The two-band model Hamiltonian in the absence of external fields is of
the form
\begin{equation}
{\cal H}=\alpha_y v_{\perp} p_\perp + \alpha_z v_z p_z + {1\over 2}\,
\beta E_g(z),
\label{1}
\end{equation}
where the $z$ axis is perpendicular to the heterojunction (it is assumed
that the growth direction is $[111]$), $E_g(z)$ stands for the position
dependent gap, $\alpha_y$, $\alpha_z$ and $\beta$ are the usual $4\times
4$ Dirac matrices, $v_\perp$ and $v_z$ have dimensions of velocity and
they are related to the Kane's matrix elements.  As usual, it is assumed
that these matrix elements are constant through the whole
heterostructure.  Since the gap and the gap centre depend only upon $z$,
the transversal momentum is a constant of motion and we can set the $y$
axis parallel to this component.  In a graded-gap structure of thickness
$L$ with band inversion we have $E_g(z)=Kz$, where
$K=(E_{gR}-E_{gL})/L$.  Here $E_{gR}>0$ ($E_{gL}<0$) is the magnitude of
the gap at $z_R$ ($-z_L$) with $L=z_R+z_L$ and $E_{gR}/E_{gL}=-z_R/z_L$.

In the two-band model there are four envelope-functions including spin,
and we arrange them in a four component spinor $F({\bf r})$, whose upper
and lower components give the coefficients of the $L_6^{-}$ and
$L_6^{+}$ parts of the wave function.  This spinor satisfies the
equation
\begin{equation}
{\cal H}\,F({\bf r}) = [E-V(z)]\,F({\bf r}),
\label{2}
\end{equation}
where $V(z)$ gives the position of the gap centre.  The way $V(z)$
changes from one material to another is not well understood, so that it
is often considered that the misalignment follows the same profile of
$E_g(z)$ \cite{Agassi1}.  Therefore $V(z)=(\lambda/2)E_g(z)$ where
$\lambda/2=(V_R-V_L)/(E_{gR}-E_{gL})$, $V_R$ and $V_L$ being the gap
centre at $z_R$ and $-z_L$, respectively.  We shall see below that the
existence of localized states requires $|\lambda|<1$, namely the gaps
must overlap (type I heterojunctions).  The same condition is found for
the existence of interface states in band inverted junctions
\cite{Korenman}.  As we have already mentioned, the momentum
perpendicular to the interface is conserved, and therefore we look for
solutions of the form $F({\bf r})=F(z)\exp(i\,{\bf r}_\perp \cdot {\bf
p}_\perp /\hbar)$ to Eq.~(\ref{2}).  The function $F(z)$ satisfies the
following equation
\begin{equation}
\left(\alpha_y v_\perp p_\perp + \alpha_z v_z p_z+{1\over 2}\,\beta
E_g(z)-E+V(z)\right)\,F(z)=0.
\label{3}
\end{equation}
A simple way to solve this equation is the Feynman-Gell-Mann ansatz
\cite{Feynman}
\begin{equation}
F(z)=\left(\alpha_y v_\perp p_\perp + \alpha_z v_z p_z+{1\over 2}\,\beta
E_g(z)+E-V(z)\right)\,\chi(z).
\label{4}
\end{equation}
After a little algebra we have
\begin{equation}
\left\{ -\,{d^2\phantom{z}\over
dz^2}+{1\over \hbar^2v_z^2} \left[ {1\over 4}
\,E_g(z)^2-[E-V(z)]^2+v_\perp^2p_\perp^2\right]+M\right\} \chi(z)=0,
\label{5}
\end{equation}
where the $4\times 4$ constant matrix $M$ is given by $M=-i(K/2\hbar
v_z) \alpha_z(\beta-\lambda)$, whose eigenvalues $\pm \mu=\pm (K/2\hbar
v_z) \sqrt{1-\lambda^2}$ are real since we take $|\lambda|<1$.  Setting
$\chi(z)=f_{\pm}(z)\phi_{\pm}$, where $\phi_{\pm}$ are the eigenvectors
corresponding to the eigenvalues $\pm \mu$, we obtain
\begin{equation}
\left[-\,{d^2\phantom{z}\over dz^2}+\mu^2z^2+{\lambda KE
\over \hbar^2v_z^2}\,z\right]\,f_{\pm}(z)=\left[{E^2-v_\perp^2p_\perp^2
\over \hbar^2v_z^2}\mp \mu\right]\,f_{\pm}(z),
\label{6}
\end{equation}
which clearly reduces to a nonrelativistic oscillator equation by
carrying out a suitable translation of the origin of coordinates (recall
that $\mu^2>0$).  Thus, inverting the various transforms needed to
arrive at (\ref{6}), it is not difficult to demonstrate that envelope
functions are simply combinations of Hermite polynomials times a
decreasing exponential factor similar to that of the harmonic
oscillator, provided that $f(z)$ vanishes at $|z|\to\infty$ in a
suitable way.  The corresponding bound levels can also be found in a
simple fashion, and they are given by

\begin{equation}
E_n^2=(1-\lambda^2)\left[ n\hbar v_zK\sqrt{1-\lambda^2} +
v_\perp^2p_\perp^2\right],
\label{7}
\end{equation}
$n$ being a positive integer. Notice that energy levels increase as the
square root of $K$ and $n$, thus being no longer equally spaced. This
behaviour is usually obtained in ``relativistic-like'' oscillator
equations \cite{Can}. As an example, let us consider the case of
selenides with symmetric band inversion, i.e., $-E_{gL}=E_{gR}\equiv
E_g$ and $V_L=V_R$ ($\lambda=0$). Typical parameters are $E_g=0.15\,$eV,
$\hbar v_z=2.76\,$eV. For $L=2z_R=2z_L=600\,$\AA\ and $p_\perp=0$ we
obtain four bound state levels with energy less than $E_g/2$:
$E_1=37.2\,$meV, $E_2=52.5\,$meV, $E_3=64.3\,$meV and $E_4=74.3\,$meV.
Hence the level spacings are $E_2-E_1=15.3\,$meV, $E_3-E_2=11.8\,$meV,
$E_4-E_3=10.0\,$meV, showing that levels are not equally spaced but
deviation from that behaviour is actually small.

To summarize, we have discussed in some detail the electronic structure
of graded-gap hetrojunctions of IV-VI compuounds presenting band
inversion.  The theoretical analysis is based on the two-band model
arising in the ${\bf k}\cdot{\bf p}$ approach which, neglecting far-band
couplings, becomes completely analogous to a Dirac-like equation with
linearly rising scalar- and electrostatic-like potentials.  The Dirac
Hamiltonian for linear potentials is exactly solvable, leading to
envelope functions that can be expressed as a sum of harmonic oscillator
functions.  Therefore, carriers are spatially localized close to $z=0$,
i.e., the plane where band inversion occurs.  However, notice that the
maximum value of envelope functions is located at $z_0=-2\lambda E/K
\sqrt{1-\lambda^2}$ since Eq.~(\ref{6}) is a harmonic oscillator
equation centered at $z_0$.  This is to be compared to interface states
lying within the fundamental gap which, in absence of external fields,
are centered at the crossing point of $L_6^{-}$ and $L_6^{+}$ bands
\cite{Korenman}.  In addition, we have found that energy levels increase
as the square root of effective coupling $K$ and the quantum number $n$.
Let us stress that the level spacing and spatial extend of envelope
functions depend on the value of $K$, and as a consequence they can be
controlled by varying the values of $E_{gL}$ and $E_{gR}$ as well as the
thickness of the structure $L$.  Finally, some words concerning the
validity of the present model are in order.  It is known that the
Hamiltonian (\ref{1}) has limitations since it neglects far-band
corrections.  Those effects can be evaluated by means of the standard
second-order perturbation theory \cite{Agassi1}, although we do not
attempt to carry out such computation here.  However, we can confidently
expect that our results are valid, at least qualitatively, since in
most selenides such corrections only cause small deviations of the
results predicted from (\ref{1}).  In fact, interface states lying
within the fundamental gap remain even if far-band terms are included in
the Hamiltonian, while only minor modifications of the dispersion
relation are observed \cite{Agassi1}.  We hope that our results may
encourage experimental effort in this field for two reasond.  First, to
validate or discard the existence of localized states in actual
band-inverted junctions other than interface states.  Second, the
feature of having oscillator-like states along with their associate
discrete spectrum may be the basis for designing new devices and
applications.

The author thanks A.\ S\'anchez for a critical reading of the
manuscript. This work is supported by DGICYT (Spain) under project
MAT95-0325.

\end{document}